\documentstyle[aps,prl]{revtex}
\begin{document}
\preprint{}
\draft
\title{Construction of a Novel Superstring in Four Dimensions}
\author{B. B. Deo}
\address{Physics Department, Utkal
University, Bhubaneswar-751004, India.} 
\maketitle
\begin{abstract}
A string in four dimensions is constructed by supplementing it with  
forty four Majorana fermions. The central charge is 26.
The fermions are grouped in
such a way that the resulting action is supersymmetric.
The super-Virasoro algebra is constructed and closed by
the use of Jacobi identity. The tachyonic ground state 
decouples from the physical states. GSO projections are necessary
for proving modular invariance and space-time supersymmetry is
shown to exist for modes of zero mass. The symmetry group of the 
model desends to the low energy group $ SU(3) \times SU(2) \times  
U(1) \times U(1) $. 

\end{abstract}

\section{Introduction}
String theory was invented \cite{bd1} as a sequel to dual
resonance models \cite{bd2} to explain the properties of 
strongly interacting particles in four dimensions. Assuming
the string to live in a background gravitational field and
demanding Weyl invariance, the Einstein equations of general
relativity could be deduced. It was believed that about these
classical solutions one can expand and find the quantum
corrections. But difficulties arose at the quantum level.
Eventhough the strong interaction amplitude obeyed crossing, 
it was no longer unitary. There were anomalies and ghosts. 
Due to these compelling reasons it was necessary for the open 
string to live in 26
dimensions \cite{bd3,bd10}. At present the most successful theory is the ten 
dimensional superstring on a Calabi-Yau manifold or an orbifold.
However, in order to realise the programme of string unification of all
the four particle interaction, one must eventually arrive at a
theory in four flat space-time dimensions, with N=1 supersymmetry
and chiral matter fields. This paper is an attempt in that direction.

A lot of research has been done in the construction of four 
dimensional strings  \cite{bd99} specially in the 
latter half of the eighties. Antoniadis
et al\cite{ia} have considered a four dimensional superstring supplemented 
by eighteen real fermions in trilinear coupling. The central charge of the
construction is 15. Chang and Kumar \cite{dc} have considered
Thirring fermions, but again with the central charge at 15. Kawai et al
\cite{hk} have considered four dimensional model in a different context
than the model proposed here. None of these makes contact with a
standard like model.

In section II, we give the details of the supersymmetric model.
Section III gives the usual quantization and super-Virasoro algebra is
deduced in the section IV. Bosonic states are constructed in Section
V. Fadeev- Popov ghosts are introduced and the
BRST charge is explicitly constucted in section VI. Ramond states 
are constructed in section VII. In section VIII  
the mass spectrum of the model and the necessary GSO projections to
eliminate the half integral spin states are introduced.  In section IX, we show that these projections are necessary to prove the modular 
invariance of the model. Space-time supersymmetry is shown to exist
for the zero mass modes in sction X. In section XI we show how the 
chain $SO(44)\to SO(11)\to SO(6)\times SO(5) \to SU(3)\times SU(2)
\times U(1) \times U(1)$ is possible in this model.

The literature on string theory is very vast and exist in most
text books on the subject. The references serve only as a guide
to elucidate the model.

\section{The Model} 
The model essentially consists of 26 vector bosons of an open (closed)
string in which there are
the four bosonic coordinates of four dimensions and there are 
fortyfour Majorana fermions representing the remaining 22 
bosonic coordinates \cite{bd11}. We divide them into four 
groups .
They are labelled by $\mu=0,1,2,3$ and
each group contains 11 fermions. These 11 fermions are again
divided into two groups, one containing six and the other five. 
For convenience, in one group we have $j=1,2,3,4,5,6$,
and in the other,
$k=1,2,3,4,5$.

The string action is
\begin{equation}
S=-\frac{1}{2 \pi} \int d^2 \sigma \left [ {\partial}_{\alpha} 
X^{\mu}{\partial}^{\alpha} X_{\mu}-i \bar{\psi}^{\mu,j}
\rho^{\alpha} {\partial}_{\alpha}\psi_{\mu,j}-i\phi^{\mu,k} 
\rho^{\alpha}{\partial}_{\alpha}\phi_{\mu,k}\right ]\;,
\end{equation}
$\rho^{\alpha}$ are the two dimensional Dirac matrices
\begin{equation}
\rho^0=\left ( \begin{array}{cc} 0 & -i\\
i & 0 \end{array} \right )\;,\;\;\;\;\;
\rho^1=\left ( \begin{array}{cc} 0 & i\\
i & 0 \end{array} \right )
\end{equation}
and obey 
\begin{equation}
\{\rho^{\alpha}, \rho^{\beta}\}=-2 \eta_{\alpha \beta}\;.
\end{equation}
String co-ordinates $X^{\mu}$ are scalars in $(\sigma ,\tau)$ space and
vectors in target space. Similarly $\psi^{\mu ,j}$ are spinors in
$(\sigma ,\tau)$ space and vectors in target space. 

In general we follow the notations and conventions of reference
\cite{bd5} whenever omitted by us. $X^{\mu}(\sigma,\tau)$ are 
the string coordinates. The $\psi$'s are the odd indexed and
$\phi$'s the even indexed Majorana fermions decomposed 
in the basis
\begin{equation}
\psi=\left (\begin{array}{c}\psi_- \\ \psi_+ \end{array}
\right )\;,
\;\;\;\mbox{and}\;\;\;\;
\phi=\left (\begin{array}{c}\phi_- \\ \phi_+ \end{array}
\right )\;.
\end{equation}
The nonvanishing commutation and anticommutations are
\begin{equation}
[\dot{X}^{\mu} (\sigma,\tau),X^{\nu}(\sigma^{\prime},\tau)]
=-i\;\delta(\sigma-\sigma^{\prime})\eta^{\mu \nu}
\end{equation}
\begin{equation}
\{\psi_A^{\mu}(\sigma,\tau), \psi_B^{\nu}(\sigma^{\prime},\tau)\}
=\pi\; \eta^{\mu \nu}\; \delta_{AB}\; \delta(\sigma-\sigma^{\prime})
\end{equation}
\begin{equation}
\{\phi_A^{\mu}(\sigma,\tau), \phi_B^{\nu}(\sigma^{\prime}, \tau)\}
=\pi\; \eta^{\mu \nu}\; \delta_{AB}\; \delta(\sigma-\sigma^{\prime})
\end{equation}

The action is invariant under infinitesimal transformations
\begin{equation}
\delta X^{\mu}=\bar \epsilon\; \left ( \sum_j \psi^{\mu,\;j}
+i\sum_k\phi^{\mu,\;k}\right )
\end{equation}
\begin{equation}
\delta\psi^{\mu,\;j}=-i \rho^{\alpha}\; \partial_{\alpha} X^{\mu}
\;\epsilon
\end{equation}
\begin{equation}
\delta \phi^{\mu,\;k}=+\rho^{\alpha}\;\partial_{\alpha}\; X^{\mu}
\;\epsilon
\end{equation}
where $\epsilon$ is an infinitesimally constant anticommuting
Majorana spinor. The commutator of the two supersymmetry
transformations gives a spatial translation, namely
\begin{equation}
[\delta_1,\delta_2]X^{\mu}=a^{\alpha}\;{\partial}_{\alpha}X^{\mu}
\end{equation}
and
\begin{equation}
[\delta_1,\delta_2]\Psi^{\mu}=a^{\alpha}{\partial}_{\alpha}
\Psi^{\mu}
\end{equation}
where 
\begin{equation}
a^{\alpha}=2i\;\bar \epsilon_1\; \rho^{\alpha}\; \epsilon_2
\end{equation}
and
\begin{equation}
\Psi^{\mu}=\sum_j\psi^{\mu,\;j}+i\sum_k\phi^{\mu,\;k}
\end{equation}
In deriving this, the Dirac equation for the spinors have been
used. The Noether super-current is
\begin{equation}
J_{\alpha}=\frac{1}{2} \rho^{\beta}\;\rho_{\alpha}
\;\Psi^{\mu}\; \partial_{\beta} X_{\mu}
\end{equation}

We now follow the standard procedure. The light cone components
of the current and energy momentum tensors are
\begin{equation}
J_+=\partial_+X_{\mu}\; \Psi^{\mu}_+
\end{equation}
\begin{equation}
J_-=\partial_-X_{\mu}\;\Psi^{\mu}_-
\end{equation}
\begin{equation}
T_{++}=\partial_+X^{\mu} \partial_+X_{\mu}+\frac{i}{2}
\psi_+^{\mu,\;j}\;\partial_+\psi_{+\mu,\;j}+\frac{i}{2}
\phi_+^{\mu,\;k}\;\partial_+\phi_{+\mu,\;k}
\end{equation}

\begin{equation}
T_{--}=\partial_-X^{\mu} \partial_-X_{\mu}+\frac{i}{2}
\psi_-^{\mu,\;j}\partial_-\psi_{-\mu,\;j}+\frac{i}{2}
\phi_-^{\mu,\;k}\partial_-\phi_{-\mu,\;k}
\end{equation} 
where $\partial_{\pm}=\frac {1}{2} (\partial_{\tau}\pm 
\partial_{\sigma})$.

To proceed further we note that in equation (8) and (14) we could have 
taken $-i$ instead of $+i$. We now introduce a phase factor
$\eta_{\phi}$ to replace $`i'$ in these equation. $\eta_{\phi}$
depends on the number $n_{\phi}$ of $\phi_{s}$ (or its quantar), in 
a given individual term. Explicitly $\eta_{\phi} = (-1)^{1/4
  n_{\phi}(n_{\phi} - 1) + \frac{1}{2}}$. $\eta_{\phi} = i$ if
$n_{\phi} = 1$ reproducing $`i'$ in the above equations. But
$\eta_{\phi}^{2} = -1$ if $n_{\phi} = 0$ where two $\phi$'s have
been contracted away and $\eta_{\phi}^{2} = 1$ if $n_{\phi} = 2$.

One now readily calculates the algebra
\begin{eqnarray}
\{ J_{+} (\sigma), J_{+}(\sigma^{\prime})\} && = \pi
\delta(\sigma - \sigma^{\prime}) T_{++} (\sigma)  \nonumber \\
\{ J_{-} (\sigma), J_{-}(\sigma^{\prime})\} && = \pi
\delta(\sigma - \sigma^{\prime}) T_{--} (\sigma) \nonumber \\
\{ J_{+} (\sigma), J_{-}(\sigma^{\prime})\} && = 0 
\end{eqnarray}

The time like components of $X^{\mu}$ are eliminated by the use of
Virasoro constraints $T_{++} = T_{--} = 0$. In view of equation
(20), we postuate that
\begin{equation}
0 = J_{+} = J_{-} = T_{++} = T_{--}
\end{equation}

$J_{+}$ is a sum of a real and imaginary term, The real term, is
further, a sum of six mutually independent $\psi^{\mu, j}$ 's and the
imaginary term, the five mutually independent $\phi^{\mu, k}$ 's. It
will be shown in Section V, that $J_{+} = 0$ constraint excludes all
the eleven time like components of $\psi$'s and $\phi$'s from the
physical space.

\section{Quantization}

As usual the theory is quantized  ($\alpha_o^{\mu} = p^{\mu}$), with
\begin{equation}
X^{\mu}=x^{\mu}+p^{\mu}\tau +i\sum_{n\neq 0}\frac{1}{n}\alpha^{\mu}_n
\exp^{-i n\tau} cos(n\sigma), \nonumber
\end{equation}
or
\begin{equation}
\partial_{\pm}X^{\mu}=\frac{1}{2}\sum_{-\infty}^{+\infty}
\alpha_n^{\mu}\; e^{-in(\tau\pm\sigma)}
\end{equation}
\begin{equation}
[\alpha_m^{\mu},\alpha_n^{\nu}]=m\; \delta_{m+n}\;
\eta^{\mu \nu}
\end{equation}

While discussing the mass spectrum, it will be more illuminating to
consider the closed string. The related additional quantas here and in
wherever occurs will be denoted by attaching a tilde. For instance
\begin{equation}
\partial_{-}X^R_{\mu}=\sum_{-\infty}^{+\infty}\alpha^{\mu}_n
e^{-2in(\sigma-\tau)}
\end{equation}
\begin{equation}
\partial_{+}X^L_{\mu}=\sum_{-\infty}^{+\infty}\tilde{\alpha}^{\mu}_n
e^{-2in(\sigma+\tau)}
\end{equation}
The transition formulas for closed strings can be easily effected. 
We consider the open string. 
We first choose the Neveu-Schwarz (NS) \cite{bd4}
boundary condition. Then the mode expansions of the fermions are
\begin{equation}
\psi_{\pm}^{\mu,j}(\sigma,\tau)=\frac{1}{\sqrt 2}
\sum_{r\in Z+\frac{1}{2}}b_r^{\mu,\;j}e^{-ir (\tau\pm\sigma)}
\end{equation}

\begin{equation}
\phi_{\pm}^{\mu,k}(\sigma,\tau)=\frac{1}{\sqrt 2}
\sum_{r\in Z+\frac{1}{2}}b_r^{\prime\,\mu,\;k}e^{-ir (\tau \pm \sigma)}
\end{equation}
\begin{equation}
\Psi_{\pm}^{\mu,j}(\sigma,\tau)=\frac{1}{\sqrt 2}
\sum_{r\in Z+\frac{1}{2}}B_re^{-ir (\tau\pm\sigma)}
\end{equation}
The sum is over all the half-integer modes.
\begin{equation}
\{b_r^{\mu,j}, b_s^{\nu, j^{\prime}}\}=\eta^{\mu \nu}\;\delta_
{j,j^{\prime}}\; \delta_{r+s}
\end{equation}
\begin{equation}
\{b_r^{\prime\,\mu,k}, b_s^{\prime\,\nu, k^{\prime}}\}=\eta^{\mu \nu}\;\delta_
{k,k^{\prime}}\;\delta_{r+s}
\end{equation}
\begin{equation}
\{B_r^{\mu}, B_s^{\nu}\}=\eta^{\mu \nu}\;
\delta_{r+s}\;.
\end{equation}

\section{Virasoro Algebra}

Virasoro generators \cite{bd6} are given by the modes of the
energy momentum tensor $T_{++}$ and Noether current $J_+$,
\begin{equation}
L_m^M=\frac{1}{\pi}\int_{-\pi}^{+\pi} d\sigma\; e^{im\sigma}\; T_{++}
\end{equation}
\begin{equation}
G_r^M=\frac{\sqrt 2}{\pi}\int_{-\pi}^{+\pi} d\sigma\; e^{ir\sigma}\; J_{+}
\end{equation}
`$ M $' stands for matter. In terms of creation and annihilation
operators
\begin{equation}
L_m^M=L_m^{(\alpha)}+L_m^{(b)}+L_m^{(b')}
\end{equation}
where
\begin{equation}
L_m^{(\alpha)}=\frac{1}{2}\sum_{n=-\infty}^{\infty}:
\alpha_{-n}\cdot\alpha_{m+n}:
\end{equation}

\begin{equation}
L_m^{(b)}=\frac{1}{2}\sum_{r=-\infty}^{\infty}
(r+\frac{1}{2}m) :b_{-r}\cdot b_{m+r}:
\end{equation}

\begin{equation}
L_m^{(b')}=\frac{1}{2}\sum_{r=-\infty}^{\infty}
(r+\frac{1}{2}m):b'_{-r}\cdot b'_{m+r}:
\end{equation}

In each case normal ordering is required. The single dot
implies the sum over all qualifying indices. 
The current generator is
\begin{equation}
G^M_r= \sum_{n = - \infty}^{\infty} \alpha_{-n} \cdot (b_{r+n} +
\eta_{\phi} b^{\prime}_{r+n}) = \sum_{n=-\infty}^{\infty} \alpha_{-n}\cdot
(b_{r+n}+i\;b'_{r+n})=\sum_{n=\infty} \alpha_n\cdot B_{r+n}
\end{equation}
Following from eqn. (33) the Virasoro algebra is
\begin{equation}
[L_m^M, L_n^M]=(m-n) L_{m+n}^M+A(m)\;\delta_{m+n}
\end{equation}
Using the relations
\begin{equation}
\left [L^M_m,\alpha_n^{\mu} \right ]=-n\alpha^{\mu}_{n+m}
\end{equation}
\begin{equation}
\left [L^M_m,B_n^{\mu} \right ]=-(n+\frac{m}{2})B^{\mu}_{n+m}
\end{equation}
we get, also
\begin{equation}
[L_m^M, G^M_r]=\left (\frac{1}{2}m-r\right ) G_{m+r}^M
\end{equation}
The anticommutator $\{G^M_r, G^M_s\}$ is obtained directly or 
by the use of the Jacobi identity
\begin{equation}
[\{G^M_r,G^M_s\},L_m^M]+\{[L_m^M,G^M_r],G^M_s\}+\{
[L_m^M,G^M_s],G^M_r\}=0
\end{equation}
which implies, consistent with equations (34) and (35),
\begin{equation}
\{G^M_r,G^M_s\}=2 L_{r+s}^M+B(r)\delta_{r+s}
\end{equation}
$A(m)$ and $B(r) $ are normal ordering anomalies. Taking the 
vacuum expectation value in the Fock ground state $|0,0\rangle $ 
with four momentum $ p^{\mu}=0$ of the commutator $[L_1,L_{-1}]$
and $[L_2, L_{-2}]$, it is easily found that
\begin{equation}
A(m)=\frac{26}{12}(m^3-m) = \frac{C}{12}(m^{3}-m)
\end{equation}
and using the Jacobi identity
\begin{mathletters}
\begin{equation}
B(r)=\frac{A(2r)}{2r}
\end{equation}
\begin{equation}
B(r) = \frac{26}{3}\left (r^2-\frac{1}{4}\right ) = \frac{C}{3} \left (
  r^2 - \frac{1}{4} \right )
\end{equation}
\end{mathletters}
The central charge $C=26$. This is what is expected.
Each bosonic coordinate contribute 1 and each fermionic
ones contribute $1/2$, so that the total central charge is +26.

For closed strings there will be another set of tilded 
generators satisfying the same algebra.

\section{Bosonic States}

A physical bosonic state $\Phi$ which should have $SO(6)\times SO(5)$ 
internal symmetry satisfies
\begin{equation}
 L_m^M\; \mid \Phi \rangle=0\;\;\;\;\;\;\;\;\;m>0
\end{equation}
\begin{equation}
 G_r^M\; \mid \Phi \rangle=0\;\;\;\;\;\;\;\;\;r>0
\end{equation}
The real and imaginary parts of $G_{r}^{M} \mid \Phi \rangle$
separately vanish for $r \;  > \; 0$. Specialising to a rest frame we
write the conditions as
\begin{equation}
\frac{1}{2} p^{0} \alpha_{m}^{0} \mid \Phi \rangle \; + \; {\rm
  (terms~quadratic~in~osillators)} \mid \Phi \rangle = 0
\end{equation}
\begin{equation}
p^{0} (b_{r}^{0,1} + \cdot \cdot \cdot + b_{r}^{0,6})\mid \Phi \rangle 
+ {\rm (terms~containing~both~\alpha~and~b~oscillators)} \mid \Phi
\rangle = 0
\end{equation}
\begin{equation}
p^{0} (b_{r}^{\prime 0,1} + \cdot \cdot \cdot + b_{r}^{0,5})\mid \Phi \rangle 
+ {\rm (terms~containing~both~\alpha~and~b^{\prime}~oscillators)} \mid \Phi
\rangle = 0
\end{equation}

The resulting states in the second terms of equations (50), (51), (52)
are different from the first terms. So we concentrate our attention for 
the vanishing of the first term only. To satisfy equation (50), as
usual the time like component $\alpha_{m}^{0}$ is excluded from
$\Phi$. As a result the second terms of equation (51) and (52) contain only
transverse osillators. $b_{r}^{0,1}$ to $b_{r}^{0,6}$  or $b_{r}^{\prime 0,1}$ to
$b_{r}^{\prime 0,5}$ are all independent anahilation operators for $r
\; > \; 0$ and there is no relation between them. Therefore $\Phi$
should not contain any of the eleven time like components
$b_{r}^{j}$'s or $b^{\prime k}_{r}$ `s, for, otherwise the equality to
zero in equations (51) and (52) cannot be achieved. Thus the vanishing 
of the energy-momentym tensor and the current excludes all the time
like components from the physical space. No negative norm state will
show up in the physical spectrum and at the same time preserve $SO(6) \times SO(5)$ internal symmetry.

Let us make a more detailed investigation to ensure that there are no
negative norm physical states. We shall do this by constructing the
zero norm states or the `null' physical states. Due to GSO condition,
which we shall study later, the physical states will be obtained by
operation of the product of even number of G's. So the lowest state
above the tachyonic state is
\[ \mid \Psi \rangle = L_{-1} \mid \chi_{1} \rangle + G_{-1/2}
G_{-1/2} \mid \chi_{2} \rangle \]
But $ G_{-1/2} G_{-1/2} = \frac{1}{2} \{  G_{-1/2}, G_{-1/2} \} =
L_{-1}$. Without loss of generality, the state is
\begin{equation}
\mid \Psi \rangle = L_{-1} \mid \tilde{\chi} \rangle
\end{equation}
This state to be physical, it must satisfy $( L_{0} - 1) \mid \Psi
\rangle = 0$ which is true if $L_{0} \mid \tilde{\chi} \rangle = 0$. The
norm $\langle \Psi \mid \Psi \rangle = \langle \tilde{\chi}\mid L_{1} L_{-1} \mid
\tilde{\chi} \rangle = 2 \langle \tilde{\chi} \mid L_{0} \mid \tilde{\chi} \rangle 
= 0$. Let us consider the next higher mass state
\[ \mid \Psi \rangle = L_{-2} \mid \chi_{1} \rangle + L_{-1}^{2} \mid
\chi_{2} \rangle + ( G_{-3/2} G_{-1/2} + \lambda G_{-1/2} G_{-3/2} )
\mid \chi_{3} \rangle + G_{-1/2} G_{-1/2}G_{-1/2}G_{-1/2} \mid
\chi_{4} \rangle  + \cdots \]
It can be shown that $G_{-3/2} G_{-1/2} \mid \tilde{\chi} \rangle = (
\beta_{1} L_{-1}^{2} + \beta_{2} L_{-2}) \mid \tilde{\chi}
\rangle$. The coefficients $\beta_{1}$ and $\beta_{2}$ can be
calculated by evaluating $\left [ L_{1}, G_{-3/2} G_{-1/2} \right ]
\mid \tilde{\chi} \rangle$ and $\left [ L_{2}, G_{-3/2} G_{-1/2} \right ]
\mid \tilde{\chi} \rangle$. $G_{-1/2}^{4}$ is proportional to
$L_{-1}^{2}$. So, in essence, we have the next excited state as
\begin{equation}
\mid \Psi \rangle = \left ( L_{-2} + \gamma L_{-1}^{2} \right ) \mid
\tilde{\chi} \rangle
\end{equation}
The condition $(L_{0} - 1)\mid\Psi\rangle=0$ is satisfied if $(L_{0} + 1) 
\mid \tilde{\chi} \rangle = 0$. Further the physical state condition
$L_{1} \mid \Psi \rangle = 0$ gives the value of $\gamma = 3/2$. The
norm is easily obtained as 
\begin{equation}
\langle \Psi \mid \Psi \rangle = \frac{1}{2} (C - 26)
\end{equation}
This is negative for $C < 26$ and vanishes for $C=26$. So the critical 
cenbtral charge is 26. It is easily checked that $L_{2} \mid \Psi
\rangle$ also vanishes for $C=26$.

To find the role of $b$ and $b^{\prime}$ modes, let us calculate the norm
of the following state with $p^{2} = 2$
\begin{equation}
(L_{-2} + 3/2 L_{-1}^{2} ) \mid 0, p \rangle = \left (
  L_{-2}^{(\alpha)} + \frac{3}{2} L_{-1}^{(\alpha)^{2}} \right ) \mid, 
0, p \rangle + \left (  L_{-2}^{(b)} + \frac{3}{2} L_{-1}^{(b)^{2}}
\right ) \mid 0, p \rangle + \left (  L_{-2}^{(b^{\prime})} +
  \frac{3}{2} L_{-1}^{(b^{\prime})^{2}} \right ) \mid 0, p \rangle
\end{equation}
The norm of the first term equal $-11$ as calculated in reference
\cite{bd5}.

Noting that $L_{-1}^{(b)} \mid 0,p\rangle = L_{-1}^{(b^{\prime}}) \mid
  0, p \rangle = 0$;  $L_{-2}^{(b)} = \frac{1}{2} b_{-3/2} \cdot
  b_{-1/2}$ and $ L_{-2}^{(b^{\prime})} = \frac{1}{2} b_{-3/2}^{\prime} \cdot 
    b_{-1/2}^{\prime}$ the norms of the second and third terms are
    $\frac{1}{4} (\delta_{\mu \mu} \delta_{j j}) = 6$ and $\frac{1}{4}
    (\delta_{\mu \mu} \delta_{k k}) = 5$ respectively. The norm of the
    state given in equation (56) is $-11 + 6 + 5 = 0$

Since $L_{1} = G_{1/2}^{2}$, $L_{1} \mid \Psi \rangle = 0$ implies
$G_{1/2} \mid \Psi \rangle  = 0$. $G_{3/2}$ can be expressed as a commutator of
$L_{1}$ and $G_{1/2}$, so that $G_{3/2} \mid \Psi \rangle =
0$. Further $L_{2} \mid \Psi \rangle = \frac{1}{2} \{ G_{3/2}, G_{1/2} 
\} \mid \Psi \rangle = 0$ and so on satisfing all the physical state conditions.
\section{Ghosts}
 
For obtaining a zero central charge so that the anomalies
cancel out and natural ghosts are isolated, Faddeev-Popov 
(FP) ghosts \cite{bd7} are introduced. The FP ghost action
is
\begin{equation}
S_{FP}=\frac{1}{\pi}\int (c^+\partial_- b_{++} + c^-
\partial_+b_{--})d^2 \sigma
\end{equation}
where the ghost fields $b$ and $c$ satisfy the anticommutator
relations
\begin{equation}
\{b_{++}(\sigma,\tau), c^+(\sigma^{\prime},\tau)\}
=2 \pi\; \delta(\sigma-\sigma^{\prime})
\end{equation}

\begin{equation}
\{b_{--}(\sigma,\tau), c^-(\sigma^{\prime},\tau)\}
=2 \pi\; \delta(\sigma-\sigma^{\prime})
\end{equation}
and are quantized with the mode expansions
\begin{equation}
c^{\pm}=\sum_{-\infty}^{\infty}c_n\; e^{-in(\tau\pm\sigma)}
\end{equation}

\begin{equation}
b_{\pm \pm}=\sum_{-\infty}^{\infty}b_n\; e^{-in(\tau\pm\sigma)}
\end{equation}
The canonical anticommutator relations for $c_n$'s and
$b_n$'s are
\begin{equation}
\{c_m,b_n\}=\delta_{m+n}
\end{equation}
\begin{equation}
\{c_m,c_n\}=\{b_m,b_n\}=0
\end{equation}

Deriving the energy momentum tensor from the action and making
the mode expansion, the Virasoro generators for the ghosts (G)
are
\begin{equation}
L_m^G=\sum_{n=-\infty}^{\infty}(m-n)\;b_{m+n}\; c_{-n}- a\; \delta_{m}
\end{equation}
where $a$ is the normal ordering constant. These generators 
satisfy the algebra
\begin{equation}
[L_m^G,L_n^G]=(m-n)\;L_{m+n}^G+A^G(m)\; \delta_{m+n}
\end{equation}
The anomaly term is deduced as before and give
\begin{equation}
A^G(m)=\frac{1}{6}(m-13m^3)+2a\;m
\end{equation}
With $a=1$, this anomaly term becomes
\begin{mathletters}
\begin{equation}
A^G(m)=-\frac{26}{12}(m^3-m)
\end{equation}
\begin{equation}
B^{G} (r) = - \frac{26}{3} \left ( r^{2} - \frac{1}{4} \right )
\end{equation}
\end{mathletters}
The central charge is $-26$ and cancels the normal ordder $A(m)$ and
$B(r)$ of the $L$ and $G$ generators. Noting that
\begin{equation}
[L^G_m, c_n]=-(2m+n)c_{n+m}
\end{equation}
it is possible to construct an equation for the generator for the current of the ghost sector,
\begin{equation}
G_r^{gh}= \sum_{p} ( \frac{p}{2} - r) c_{-p} G^{gh}_{p+r}
\end{equation}
such that 
\begin{equation}
[L_m^G,G_r^{gh}]=(m/2-r)G^{gh}_{m+r}
\end{equation}

Let us examine the possibility of an expression for the current as
\begin{equation}
G_{r}^{gh} = \sum_{n} b_{n+r} c_{-n}
\end{equation}
$n$ takes only integral values whereas $r$ is half integral. $b_{n+r}$ 
is therefore outside the ghost sector. It has a conformal dimension of 
$5/2$ so that $G_{r}^{gh}$ has the correct conformal dimension
$3/2$. The commutator
\begin{equation}
\left [ L_{m}^{g}, b_{n+r} \right ] = \left ( \frac{3}{2} m - n - r
\right ) b_{n+r+m}
\end{equation}
is easily evaluated demanding that equation (70) be satisfied. From
Jacobi identity (65)
\begin{equation}
\{ G_{r}^{gh}, G_{s}^{gh} \} = 2 L_{r+s}^{G} + \delta_{r+s} B^{G}(r)
\end{equation}
The total current generator is 
\begin{equation}
G^r=G_r^M+G^{gh}_r
\end{equation}
thus we have the anomaly free Super Virasoro algebra,

\begin{equation}
[L_m,L_n]=(m-n)L_{m+n}
\end{equation}

\begin{equation}
[L_m,G_r]=(m/2-r)G_{r+m}
\end{equation}

\begin{equation}
[G_r,G_S]=2L_{r+s}
\end{equation}
 Thus from the usual conformal field theory we have 
obtained the algebra of a superconformal 
field 
theory. This is the novelty of the present formulation.
The BRST \cite{bd8} charge operator is
\begin{equation}
Q_{BRST}=\sum_{-\infty}^{\infty}L_{-m}^M\;c_m -\frac{1}{2}
\sum_{-\infty}^{\infty}(m-n) :c_{-m}\; c_{-n}\; b_{m+n} :
-a\; c_0
\end{equation}
and is nilpotent for $a=1$. The physical states are such
that $Q_{BRST}\;|phys\rangle=0$.

\section{Fermionic States}

The above deductions can be repeated for Ramond sector~\cite{ramond}. 
We write the main equations. The mode expansion for the fermions are
\begin{equation}
\psi_{\pm}^{\mu,j}(\sigma ,\tau) = \frac{1}{\sqrt 2}
\sum_{-\infty}^{\infty} d_m^{\mu,j} e^{-im(\tau\pm\sigma)}
\end{equation}
\begin{equation}
\phi_{\pm}^{\mu,j}(\sigma ,\tau) = \frac{1}{\sqrt 2}
\sum_{-\infty}^{\infty} d_m^{'\mu,j} e^{-im(\tau\pm\sigma)}
\end{equation}

The generators of the Virasoro operators are
\begin{equation}
L_m^M = L_m^{(\alpha)} + L_m^{(d)} + L_m^{(d')}
\end{equation}
\begin{equation}
L_m^{(d)} = \frac{1}{2}\sum_{n=-\infty}^{\infty}(n+\frac{1}{2} m)
: d_{-n}\cdot d_{m+n} :
\end{equation}
\begin{equation}
L_m^{(d')} = \frac{1}{2}\sum_{n=-\infty}^{\infty}(n+\frac{1}{2} m)
: d'_{-n}\cdot d'_{m+n} :
\end{equation}
and the fermionic current generator is
\begin{equation}
F_m^M = \sum_{n=-\infty}^{\infty}\alpha_{-n}\cdot 
(d_{n+m} + i d'_{n+m}) = \sum_{-\infty}^{\infty}\alpha_{-n}\cdot D_{n+m}
\end{equation}
The Ramond sector Virasoro algebra is the 
same as the NS-sector with the replacement of
G's by F's. It is necessary to define $L_o$ suitably to keep the anomaly 
equations
 the same~\cite{bd5}.

In this Ramond sector, a physical state $\mid \Phi \rangle$ should
satisfy
\begin{equation}
F_{n} \mid \Phi \rangle = L_{n} \mid \Phi \rangle = 0 \; \; \; {\rm
  for} \; \; \; n>0
\end{equation}

The normal order anomaly constant in the anticommutables of the Ramond 
current generators has to be evaluated with care, beacuse the
defination of $F_{0}$ does not have a normal ordering ambiguity. So
$F_{0}^{2} = L_{0}$. Using commutation relation (43) with $G$
replaced by $F$ and the Jacobi Identity we get
\[ \{ F_{r}, F_{-r} \} = \frac{2}{r} \{ [ L_{r}, F_{0} ], F_{-r}
\} = 2 L_{0} + \frac{4}{r} A(r) \]
So
\begin{equation}
B(r) = \frac{4}{r} A(r)
\end{equation}
\begin{equation}
B(r) = \frac{C}{3} (r^{2} - 1), \; \; \; \; r \neq 0
\end{equation}
A physical state in the fermionic sector satisfies
\begin{equation}
( L_{0} - 1 ) \mid \Psi \rangle = 0
\end{equation}
It follows that
\[ (F^{2}_{0} - 1 ) \mid \Psi \rangle  = (F_{0} - 1) (F_{0} + 1) \mid \Psi
\rangle = 0 \]

The construction of `null' physical states becomes much simpler
beacuse all $F_{-m}$ terms can be assigned to $L_{-m}$ terms by the
commution ration $F_{-m} = 2 [ F_{0}, L_{-m} ]/m$ and $F_{0}$ has
eigen values which are roots of eigen values of $L_{0}$ acting on the
generic states or states constructed out of the quadratic states. Thus 
the zero mass null physical state with $L_{0} \mid \tilde{\chi}
\rangle = F_{0}^{2} \mid \tilde{\chi} \rangle = 0$ is simply
\begin{equation}
\mid \Psi \rangle = L_{-1} \mid \tilde{\chi} \rangle
\end{equation}
with $L_{1} \mid \Psi \rangle = F_{1} \mid \Psi \rangle = 0$. The next
 excited state with $(L_{0} + 1 \tilde{\chi} \rangle )$ becomes the same as in the bosonic
  sector. Obtained from the condition $L_{1} \mid \Psi \rangle = 0$,
\[ \mid \Psi \rangle = (L_{-2} + \frac{3}{2} L_{-1}^{2}) \mid
\tilde{\chi} \rangle \]
The norm $\langle \Psi  \mid \Psi \rangle = (C - 26)/2$ and vanishes
for $C=26$. It is easy to check that all physical state conditions are 
satisfied. $F_{1} \mid \Psi \rangle = 2$ $[ L_{1}, F_{0} ] \mid \Psi \rangle = 0$
since $L_{1} \mid \Psi \rangle = 0$ and $F_{0} \mid \Psi \rangle =
\mid \Psi \rangle $ $L_{2} \mid \Psi \rangle = F_{1} F_{1} \mid \Psi
\rangle = 0$ and $F_{2} \mid \Psi \rangle = [ L_{2}, F_{0} ] \mid \Psi 
\rangle = 0$. For $C=26$, there are no negative norm states in the
Ramond sector as well.

The ghose curreent in the Ramond sector satisfies the equation 
\begin{equation}
F_{m}^{gh} = \sum_{p} ( \frac{p}{2} -m ) c_{-p} F_{m+p}^{gh}
\end{equation}
so that
\begin{equation}
\left [ L_{m}^{G}, F_{n}^{gh} \right ] = \left ( \frac{m}{2} -n \right 
) F_{m+n}^{gh}
\end{equation}
we can construct $F_{0}^{gh}$ with the help of an anti commuting
object $\Gamma_{n}$ which satisfy 
\begin{equation}
\{ \Gamma_{n}, \Gamma_{m} \} = 2 \delta_{m,n}
\end{equation}
It is important to write $L_{0}^{G}$ in terms of positive integrals as 
\begin{equation}
L_{0}^{G} = \sum_{n=1}^{\infty} n (b_{-n} c_{n} + c_{-n} b_{n} )
\end{equation}
It is found that
\begin{equation}
F_{0}^{gh} = \sum_{n=1}^{\infty} \sqrt{n} \Gamma_{n} ( b_{-n} c_{n} +
c_{-n} b_{n} )
\end{equation}
All other F's can be constructed by the use of the equations of super
Virasoro algebra.

From equation (67) and (86), the ghost current anomaly constant is
$B^{G} (r) = - \frac{26}{3} (r^{2} - 1)$ and cancels out the $B(r)$ of
equation(87). The total current anomalies in both the sectors
vanish.

\section{The Mass Spectrum}

The ghosts are not coupled to the physical states.
Therefore the latter must be of the form (up to null state)\cite{bd9}.
\begin{equation}
|\{n\}\; p\rangle_M \otimes\; c_1|0\rangle_G\label{eq64}
\end{equation}
$|\{n\}\; p\rangle_M$ denotes the occupation numbers and momentum of 
the physical matter states. The operator $c_1$ lowers the
energy of the state by one unit and is necessary for BRST
invariance. The ghost excitation is responsible for lowering
the ground state energy which produces the tachyon. 
\begin{equation}
(L_0^M-1)\;|phys\rangle=0
\end{equation}
Therefore, the mass shell condition is
\begin{equation}
\alpha^{\prime} M^2 = N^B+N^F-1
\end{equation}
where 
\begin{equation}
N^B=\sum_{m=1}^{\infty}\alpha_{-m}\; \alpha_m
\end{equation}
or
\begin{equation}
N^F=\sum_{r=1/2}^{\infty}r\;(b_{-r}\;\cdot  
b_r+b'_{-r}\;\cdot b'_r)\,\,\,\,\, (NS).
\end{equation}

Due to the presence of Ramond and Neveu-Schwartz sectors with periodic and 
anti-periodic 
boundary conditions, we can effect a 
GSO projection ~\cite{bd10} on the mass 
spectrum on the NSR model~\cite{bd13}. Here the
projection should refer to the unprimed and primed quantas separately. 
The desired projection is
\begin{equation}
G = \frac{1}{4} (1 + (-1)^F)(1 + (-1)^{F'})
\end{equation}
where $ F = \sum b_{-r}\cdot b_r$;    $F' =\sum b'_{-r}\cdot b'_r$ .
This eliminates the half integral values 
from the mass spectrum by choosing G=1.

For closed strings we have a similar separation 
as in Eq.\ (\ref{eq64}), namely
 the left-handed states will be in the form 
\begin{equation}
|\{ \tilde{n} \}\; p\rangle_M \otimes\; \tilde{c}_1|0\rangle_G
\end{equation}
The mass spectrum can be written as 
\begin{equation}
\frac{1}{2}\alpha ' M^2 = N + \tilde{N} -1 -\tilde{1}
\end{equation}

\section{Modular Invariance}

The GSO projection is necessary for the modular invariance of the
theory. We follow the notation of Seiberg and Witten \cite{bd98}.
Following Kaku \cite{bd13}, the spin structure $\chi (--,\tau)$
for a single fermion is given by
\begin{equation}
\chi(--,\tau) =q^{-1/24} \;Tr\; q^{2\sum_n  n\;\psi_{-n} \psi_n}
=q^{-1/24} \prod_{n=1}^\infty (1+ q^{2n-1}) = 
\sqrt{\frac{\Theta_3(\tau)}{\eta(\tau)}}\;,
\end{equation}
where $\Theta$'s will be the Jacobi Theta functions $\Theta (\theta, \tau)$
\cite{bd98}, $q=e^{i\pi\tau}$ and 
$ \eta (\tau) (2\pi) = \Theta_1^{\prime 1/3}(\tau)$. 
The path integral functions
of Seiberg and Witten for the twenty four unprimed oscillators are
\begin{equation}
A((--),\tau) = (\Theta_3 (\tau)/ \eta(\tau))^{12}\;,
\end{equation}
This is normalised to one.
\begin{equation}
A((+-),\tau) = A((--),\frac{\tau}{1+\tau}) = 
-(\Theta_2 (\tau)/ \eta(\tau))^{12}\;,
\end{equation}
\begin{equation}
A((-+),\tau) = A(+-, -\frac{1}{\tau}) =- (\Theta_4(\tau)/\eta(\tau))^{12}\;,
\end{equation}
\begin{equation}
A((++).\tau) =0\;.
\end{equation}
It is easily checked that the sum
\begin{equation}
A(\tau) = (\Theta_3 (\tau)/\eta(\tau))^{12} - 
(\Theta_2 (\tau)/\eta(\tau))^{12}-(\Theta_4(\tau)/\eta(\tau))^{12}
\end{equation}
is modular invariant, using the properties of the theta functions given in
\cite{bd98}.

For the twenty primed oscillators it is not so straightforward because
of the ambiguity of fractional powers of unity. If we prescribe a 
normalization $1=1^{1/2}= \sqrt{e^{2i\pi}}$, then
\begin{equation}
A^\prime ((--),\tau) = (\Theta(\tau)/\eta(\tau))^{10}\;,
\end{equation}
\begin{equation}
A^\prime((+-),\tau) = \sqrt{e^{i\pi}}(\Theta_2(\tau)/\eta(\tau))^{10}\;,
\end{equation}
\begin{equation}
A^\prime((-+),\tau)=\sqrt{e^{i\pi}} (\Theta_4(\tau)/\eta(\tau))^{10}\;,
\end{equation}
\begin{equation}
A^\prime((++),\tau) =0\;.
\end{equation}
The sum $(\Theta_3^{10} (\tau) +\sqrt{e^{i\pi}}\Theta_2^{10} (\tau)
+\sqrt{e^{i\pi}} \Theta_4^{10}(\tau))/\eta^{10}(\tau)$ is also 
modular invariant upto the factor of cube root and fractional roots
of unity. The sum of the modulii is, of course, modular invariant.
It is easy to construct the modular invariant partition function for the
two physical bosons, namely
\begin{equation}
{\cal P}_B(\tau) =(Im \;\tau)^{-2}\Delta^{-2}(\tau){\bar \Delta}^{-2}(\tau)\;,
\end{equation}
in four dimensions \cite{bd96}. For normal ordering constant is $-$1 in
$q^{2(L_0-1)}$ , $-$2/12 comes from the bosons 
and $-$44/24 comes from the
fermions adding to  $-$1 for both NS and R sectors.

\section{Space-time Supersymmetry}

Let us construct the vector state of zero mass as
\begin{equation}
\alpha_1^\mu |0 \rangle e_\mu (p)\;.
\end{equation} 
The condition ${p^\mu}^2 =0$ and $L_1|\phi\rangle =0$ gives
the equation of motion and Lorentz condition,
\begin{equation}
p^2 e_\mu =0 \;\;\;\;  and \;\;\; p^\mu e_\mu =0
\end{equation}
The Ramond fermion give a state
\begin{equation}
D_{-1}^\mu |0\rangle\gamma^\mu\psi(p)\;.
\end{equation}
$\psi(p)$ is a dirac spinor of zero mass, $\gamma^\mu$ are the usual
Dirac Gamma matrices. The condition $F_1|\phi\rangle =0$ gives the Dirac
equation
\begin{equation}
\gamma \cdot p \;\psi(p) =0\;.
\end{equation}
The bosonic degrees of freedom match the fermionic degrees of freedom.

For closed strings a similar analysis can be made with $\alpha_{-1}^\mu
\tilde \alpha_{-1}^\nu |0\rangle F^{\mu\nu}$ and $ \alpha_{-1}^\mu
\tilde D_{-1}^\nu |0\rangle \gamma^\nu \psi_\mu$ following the
details given by G.S.O., we get a massless antisymmetric tensor ,
massless scalar and a massless symmetric tensor (spin 2) adding upto
4 degrees of freedom. The fermionic sector contains a massless spin
3/2 and a massless 1/2 adding upto a total of four degrees of freedom
as has been shown by G.S.O. for the four dimensional case.

\section{Approach to Standard Model}

The main motivation of constructing this theory is to see if the
internal symmetry group makes contact with the Standard Model which
explains all experimental data upto date with a high degree of accuracy.
The internal symmetry of the group was $SO(44)$. We have divided them
into four $SO(11)$'s and further subdivided to $SO(6)$ and $SO(5)$.
According to Slansky \cite{bd12} we can have the breaking
\begin{equation}
SO(11)\supset SO(6)\times SO(5) \supset G^{st}
\end{equation}
where
\begin{equation}  
G^{st} = SU_C(3) \times SU_L(2) \times U_Y(1)\times U^\prime(1)\;.
\end{equation} 
This has the content of the Standard Model with an additional gauge boson.
The 15 gauge fields of $SO(6)$ and 10 gauge fields
of $SO(5)$ are contained in  the states $H_{jj^\prime}^{\mu\nu} (p)$ and 
$H_{kk^\prime}^{\prime \mu\nu}(p)$ which are massless $(p_\mu^2 =0)$ with 
the structure
\begin{equation}
b_{-1/2}^{\mu, j} b_{-1/2}^{\nu,j^\prime} |0\rangle H_{jj^\prime}^{\mu\nu}(p)
\end{equation}
and
\begin{equation}
b_{-1/2}^{\prime \mu, k} b_{-1/2}^{\prime \nu,k^\prime} |0\rangle 
H_{kk^\prime}^{\prime \mu\nu}(p)
\end{equation}
From the condition $F_1|\phi \rangle =0$ we have $p^\mu H^{\mu\nu}
=p^\nu H^{\mu\nu}= p^\mu H^{\prime \mu\nu} = p^\nu H^{\prime \mu\nu} =0$.
There are many vectors and scalars. The vectors are like
$p^\mu A_{jj^\prime}^\nu + p^\nu A_{jj^\prime}^\mu$ and  
$p^\mu A_{kk^\prime}^{\prime \nu} + p^\nu A_{kk^\prime}^{\prime \mu}$.
The potentials satifies the equation $\Box A^{\nu}_{jj^{\prime}}  0$,
$p^{\nu} A^{\nu}_{jj} = 0$ and $\Box A^{\nu}_{k k^{\prime}} = 0$,
$p^{\nu} A^{\nu}_{k k^{\prime}} = 0$. The $(i, j^{\prime})$ and $(k,
k^{\prime})$ antisymmetric parts are the
$\underline{15}$ and $\underline{10}$ of $SO(6)$ and $SO(5)$ respectively.

There are also the massless Ramond spinors $d_{-1}^{\mu,j} |0\rangle \gamma^\mu
\psi_j $ six in number, and $d_{-1}^{\prime \mu,k} |0\rangle \gamma^\mu
\psi_k^\prime $, five in number. So in the product space there are 30 chiral 
spinors which exactly correspond to the 15 chiral fermions and 15 chiral
antifermions in one generation of the Standard Model. The particles,
$e_R, u_R, d_R, (\nu,e)_L$ and $ (u, d)_L$ and their antiparticles can be 
assigned the usual quantum numbers under $G^{st}$. 
The $G^{st}$ with fifteen fermions is exactly
space-time super-symmetric has been proven 
in reference \cite{bd22} recently. 

The fifteen Ramond fermions ca be placed in five groups each
containing three particles

\begin{equation}
\left ( \begin{array}{c} u^{B}\\ u^{y}\\ u^{R} \end{array} \right )_{L} \;\;\; \left ( \begin{array}{c} d^{B}\\ d^{y} \\ d^{R}
  \end{array} \right )_{L} \; \; \; \left ( \begin{array}{c} u^{B} \\
    u^{y} \\ u^{R} \end{array} \right )_{R} \; \; \; \left ( \begin{array}{c}
  d^{B}\\ d^{y} \\ d^{R} \end{array} \right )_{R} \; \; \; \left (
\begin{array}{c} e_{R}\\ e_{L} \\ \nu_{L} \end{array} \right )
\end{equation}

with the help of the diagonal $SU(3)$ Gallmann matrices $\lambda^{8}$
and $\lambda^{3}$ and Isospin $I_{3}$, one can construct a matrix
operator $Q_{c} = I_{3} + \frac{1}{4} ( \sqrt{3} \lambda^{8} -
\lambda^{3})$ which has three eigen values for the first four sets and 
zero eigen values for the courless leptons.

It is of interest to note that Higgs boson is classically
tachyonic. We cannot show that there are three generations of
fermions. Perhaps it 
is related to the number of ways the 30 chiral fermions can be placed under
$G^{st}$. There are many aspects which are not clear to us, 
but we have made out a case to study this model further to relate
Gravity and Standard Model through string theory with a unique vacuum.

We have profited from discussions with Dr. J.  Maharana and Dr. S. Mahapatra.
The Library, Computer and other facilities extended by Institute
of Physics, Bhubaneswar is thankfully acknowledged.

\end{document}